\documentclass[12pt,a4paper]{article}
\usepackage{epsfig,graphicx}
\newcommand{\be}{\begin{equation}}
\newcommand{\ee}{\end{equation}}
\newcommand{\bea}{\begin{eqnarray}}
\newcommand{\eea}{\end{eqnarray}}
\newcommand{\beas}{\begin{eqnarray*}}
\newcommand{\eeas}{\end{eqnarray*}}
\newcommand{\ba}{\begin{array}}
\newcommand{\ea}{\end{array}}
\newcommand{\nn}{\nonumber}

\newcommand{\bt}{\begin{table}}

\newcommand{\al}{\alpha}

\begin{document}

\title{\bf
Towards  the rotating scalar-vacuum black holes
}
\author{
Yu.~F.~Pirogov 
\\
{\small \em
Theory Division, Institute for High Energy Physics,  Protvino, 
Moscow Region, Russia}
}
\date{}
\maketitle

\begin{abstract}
\noindent
The problem of merging  in black holes  the rotation  and a massless  scalar
field is treated. It is argued in the comment  that  the Kerr-like and
Demianski-like metrics 
in General Relativity with a free massless scalar field or in the free
Brans-Dicke theory, claimed so far in the literature as an application of the
Newman-Janis algorithm, prove to be invalid.  Finding the missing stationary
axisymmetric scalar-vacuum solution(s), containing the Kerr one,  is still 
urgent.
\end{abstract}

A (near) massless scalar field may be an important ingredient of the nature,
having conceivably a lot of facets.
In particular, such a field may  drastically change the structure of the 
event horizon of the  black holes (BHs)~\cite{Newman2}, \cite{Agnese1}. 
The rotation of BHs, described by the Kerr metric~\cite{Kerr},  strongly
influencing their structure as well, merging the
two effects gets urgent. 
Newman and Janis  originally proposed  an algorithm (NJA)~\cite{Newman} for
generating a new 
stationary axisymmetric metric from a known
static spherically-symmetric one.
In the context of the BH event horizon,  the algorithm
was used in  Ref.~\cite{Agnese} (Sec.~IV) to claim   a
scalar-modified Kerr metric  in General Relativity (GR)  with a
free massless scalar field.
Previously in Ref.~\cite{Krori1},  using NJA 
a so-called   Kerr-like metric  was claimed in the  free Brans-Dicke (BD)
theory. The  generalization  to a Demianski-like metric, the latter
incorporating  the Kerr-like one, was then claimed in~\cite{Krori2}. 
In this comment  we  argue  that the above-mentioned  claims are  
incorrect,\footnote{So, applying the claimed metrics  to
observations~\cite{Nandi} is to be  taken with
care.} with the missing solutions being  still  urgent.

For simplicity, it suffices to  restrict the explicit consideration  by the
Kerr-like case in the null coordinates in GR  or in  the Einstein frame of the
free BD theory.\footnote{It goes without saying that in the 
BD theory the invalidity statement fulfills, if any,   irrespective of   the
Einstein or Jordan frames. The same, of course, concerns the transformation 
from the null coordinates  to the Boyer-Lindquist ones.}
Applying NJA  one can bring the claimed  Kerr-like metric 
in the standard notation to the form:
\bea\label{putative}
ds^2&=&(P/Q)^\alpha (du+F_{,\theta} \sin\theta d\varphi)^2\nn \\
&& +\ 2 (du+F_{,\theta} \sin\theta d\varphi)(dr-F_{,\theta} \sin\theta
d\varphi)\nn\\
&&- \  (P/Q)^{1-\alpha} Q (d\theta^2  +\sin^2\theta d\varphi^2),\\
P&=& r^2-2r_0 r +F^2,\nn\\
Q&=&r^2+F^2,\nn\\
F&=& a \cos\theta,\nn
\eea
where $r_0$, $a$ and $\alpha$ are some parameters,\footnote{Cf., e,g.,
Eq.~(14) in Ref.~\cite{Krori1} for the Jordan frame of the BD theory, with 
transition to the Einstein frame achieved  by  the BD parameter
substitutions: $\eta\to \alpha$ and  $\xi\to 1-\al$.  The
same  follows from Eq.~(35) in Ref.~\cite{Agnese} for GR (prior to 
transformation to the  Boyer-Lindquist coordinates), with the
(other) parameter substations: $\eta  \to r_0$ and  $m/\eta \to \al$.} 
setting, respectively, the BH ``bare'' mass, specific
angular momentum and the index of the  scalar field.\footnote{At that, the
BH gravitating mass  defined by the Newtonian limit  is  $m=\al
r_0$, with the maximal $\al=1$ corresponding to the zero field.}
At $a=0$ and arbitrary $\al$, Eq.~(\ref{putative}) reproduces the exact
spherically-symmetric solution with a free massless scalar field (omitted here).
At  $\alpha =1$ and arbitrary  $a$, it  gives the concise 
description of the (scalar-less) Kerr metric.   Nevertheless at  the
arbitrary $a$ and $\al$,   Eq.~(\ref{putative})
ceases, generally,   to be correct, contrary to what one might superficially
expect.

Indeed, the result of applying NJA  is a priori  ambiguous  and needs
ultimate verification (if any)  through the field equations (FEs). In the case
at hand,  the  putative metric should satisfy the
vacuum  gravity EEs, $R_{\mu\nu}=0$, except 
for the $rr$, $r\theta$ and $\theta\theta$-elements, which are influenced,
generally,  by a scalar field $\phi(r, \theta)$.  In particular,  accounting for
(\ref{putative}) we get the $ur$-component of the Ricci tensor as follows: 
\be
R_{ur}=2\frac{F^2}{Q^2}\bigg(\frac{P}{Q}\bigg)^{\al-2}
\Bigg(\bigg(\frac{P}{Q}
\bigg)^{2\al}- \bigg(\frac{P}{Q}\bigg)^\al-\al\frac{P}{Q}\bigg(
\frac{P}{Q}-1\bigg)\Bigg).
\ee
For this to be zero identically,  there should fulfill either  $F=0$, implying 
$a=0$, or
\be
\bigg(\frac{P}{Q}\bigg)^\al= \frac{1}{2} +  \Bigg( \frac{1}{4}  +   \al 
\frac{P}{Q}
\bigg( \frac{P}{Q}-1\bigg)  \Bigg)^{1/2},
\ee
implying $\al=1$.\footnote{There  are also two 
marginal cases: either $P/Q=1$, implying $r_0=0$, or   $\al=0$.  The
first, $\al$-independent case may  ultimately be reduced  to the  scalar-less
flat one.
The second case proves to imply, in turn, either $r_0=0$, reducing to the
previous case, or $a=0$, reducing to the particular spherically-symmetric
scalar-vacuum case.} This signifies  that the  Kerr-like
metric (\ref{putative})  at the arbitrary $a$ and $\al$ is invalid, despite
the claims on the contrary~\cite{Agnese},
\cite{Krori1}.\footnote{Note also that  Refs.~\cite{Agnese} and \cite{Krori1}
claim the principally different scalar fields under the same  metric (after
transforming the BD theory to the Einstein frame).
However,  the  invalidity  statement is  independent of  the
particular  scalar field. }  The same applies to  the claimed Demianski-like
metric~\cite{Krori2} containing, in particular,  the Kerr-like one.
Thus  finding the correct stationary axisymmetric  
scalar-vacuum solution(s), containing the Kerr one,  remains still  an urgent
problem.

\paragraph{Acknowledgment} Thanks are due to I.Yu.~Polev for assistance with
the computer algebra calculations.

\end{document}